\documentclass[aps,prb,twocolumn,showpacs,ams]{revtex4-1}

\usepackage{graphicx}
\usepackage{bm}
\usepackage{amssymb} 
\usepackage{amsmath}

\begin{document}

\title{Coherent control of indirect excitonic qubits in optically driven quantum dot molecules}

\author{Juan E. Rolon}
\email{rolon@phy.ohiou.edu}
\affiliation{Department of Physics
and Astronomy and Nanoscale and Quantum Phenomena Institute, Ohio
University, Athens, Ohio 45701-2979}

\author{Sergio E. Ulloa}
\affiliation{Department of Physics and Astronomy and Nanoscale and
Quantum Phenomena Institute, Ohio University, Athens, Ohio
45701-2979}

\date{\today}

\begin{abstract}
We propose an optoelectronic scheme to define and manipulate an indirect neutral exciton qubit within a quantum dot molecule. We demonstrate coherent dynamics of indirect excitons resilient against decoherence effects, including direct exciton spontaneous recombination. For molecules with large interdot separation, the exciton dressed spectrum yields an often overlooked avoided crossing between spatially indirect exciton states. Effective two level system Hamiltonians are extracted by Feshbach projection over the multilevel exciton configurations. An adiabatic manipulation of the qubit states is devised using time dependent electric field sweeps. The exciton dynamics yields the necessary conditions for qubit initialization and near unitary rotations in the picosecond time scale, driven by the system internal dynamics. Despite the strong influence of laser excitation, charge tunneling, and interdot dipole-dipole interactions, the effective relaxation time of indirect excitons is much longer than the direct exciton spontaneous recombination time, rendering indirect excitons as potential elemental qubits in more complex schemes.
\end{abstract}

\maketitle

\newcommand{\be}   {\begin{equation}}
\newcommand{\ee}   {\end{equation}}
\newcommand{\ba}   {\begin{eqnarray}}
\newcommand{\ea}   {\end{eqnarray}}
\newcommand{\maxim}   {\mbox{\scriptsize max}}
\newcommand{\reduced}   {\mbox{\scriptsize red}}
\newcommand{\pol}   {\mbox{\scriptsize pol}}
\newcommand{\imp}{\mbox{\scriptsize imp}}

\section{Introduction}
\label{sec: Intro}

Semiconductor quantum dot molecules (QDMs) are potential building blocks for solid state quantum computation architectures.\cite{Bayer} These devices have remarkable electronic and optical properties, arising from their molecular exciton spectrum.\cite{Krenner, Bester} Moreover, the optical response of QDMs is highly tunable using external electric and magnetic fields. Quantum optics techniques have demonstrated coherent phenomena in QDMs, such as Rabi oscillations and level anticrossing of excitonic dressed states.\cite{Sham,Rabi, Stinaff, Scheibner} Strong localization of charge and spin in these structures permits multiple ways of harnessing charge and spins qubits, all limited by uncontrolled interactions within the molecule environment.\cite{Control} On one hand, spin qubits in self-assembled quantum dot structures have received a great deal of attention due to the large coherence time of spins localized in quantum dots, limited mostly by their hyperfine interactions with the QDM nuclear spin reservoir.\cite{NuclearBath} In particular, optical spin initialization and non-destructive measurements have already been implemented in QDMs.\cite{Initialization} On the other hand, charge qubits have typically shorter decoherence times, limited by spontaneous exciton recombination and electron-phonon interactions.\cite{Spontaneous} The latter limitation can be largely suppressed at very low temperatures while the former is more subtle. For neutral spatially direct exciton qubits, with logic states typically embodied in the presence or absence of an exciton in a single QD, spontaneous recombination is fast ($\sim 1$ns) and highly detrimental, due to the large direct exciton oscillator strength.

In this work we investigate the exciton dynamics of optically driven and electrically gated QDMs coupled by charge tunneling and F\"{o}rster  energy transfer (FRET).\cite{FRET, FRET_Exp_Self, Govorov} We argue, theoretically and numerically, that an exciton dressed qubit,\cite{dressqubits, solidqubits, nielsenchuang} with logical states constituted by neutral indirect excitons, can be effectively extracted from the QDM exciton dressed spectrum. Our work indicates that a control scheme is devisable using external electric field sweeps. By these means, the qubit can be initialized and rotated multiple times with high fidelity, well beyond the spontaneous exciton recombination time scale. Furthermore, we devise a read out scheme, using an adiabatic population transfer of the output indirect state into an auxiliary direct exciton state. This opens the possibility of a realistic realization of a neutral exciton qubit with enhanced characteristics and subsequent coherent manipulation using optical and electrical means. For typical QDM structures, we find that one can anticipate relaxation/decoherence times of at least two-orders of magnitude larger than spontaneous recombination times. Further separating the dots in the QDM could give rise to even longer coherence times, making them suitable for implementing complex multiqubit architectures.

In Sec.\  \ref{sec:Model} we introduce a realistic model for the QDM, that takes into account all relevant electron-hole states, and processes at the relevant energies. An excitonic dressed spectrum and population bias map is introduced in Sec.\ \ref{spectrum} and used to indicate the different molecular resonances (level anticrossings). In Sec.\  \ref{extraction} we employ a Feshbach projection formalism,\cite{Feshbach} adiabatically eliminating selected exciton transitions and extracting an effective Hamiltonian describing locally the relevant level anticrossings. In particular, we demonstrate that adiabatic elimination of the direct exciton transitions leads to an effective qubit subspace consisting of two long lived indirect excitons.
In Sec.\  \ref{3levels} we describe the qubit dynamics in two regimes, strong vacuum-indirect exciton coupling (qubit initialization) and strong coupling among two molecular indirect excitons (qubit rotation). We discuss the role of FRET and biexciton states in Sec.\ \ref{FRET-Biexciton} and find that the qubit subspace is resilient against their detrimental effects as well as to small corrections to the laser detuning. Sec.\ \ref{control} introduces an applied adiabatic bias ramp that implements the initialization, rotation and readout of the qubit logical states. It is demonstrated that the qubit can be initialized and rotated  with near unity fidelity within a picosecond scale. Finally, in Sec.\ \ref{dissipation}, we develop a method to extract effective decay rates of the molecular indirect excitons during initialization and rotation regimes, which results in relaxation times many orders of magnitude larger than the manipulation times.

\section{Model}
\label{sec:Model}

The QDM consists of two vertically stacked non-identical ``top" (T) and ``bottom" (B) quantum dots. The dots are separated by a barrier of thickness $d$ and subject to an applied axial electric field $F$ that results from the application of a top gate voltage. This is realized by placing the QDM in a $n$-$i$ Schottky junction.\cite{Krenner} The QDM is pumped by a broad square laser pulse of frequency $\omega$, which may excite different nearby exciton levels. The pulse duration is long enough, typically $\geq 1$ps, to capture several amplitude oscillations of the excitonic populations.

\begin{figure}[htb]
\includegraphics[totalheight=0.6\columnwidth,width=1.0\columnwidth]{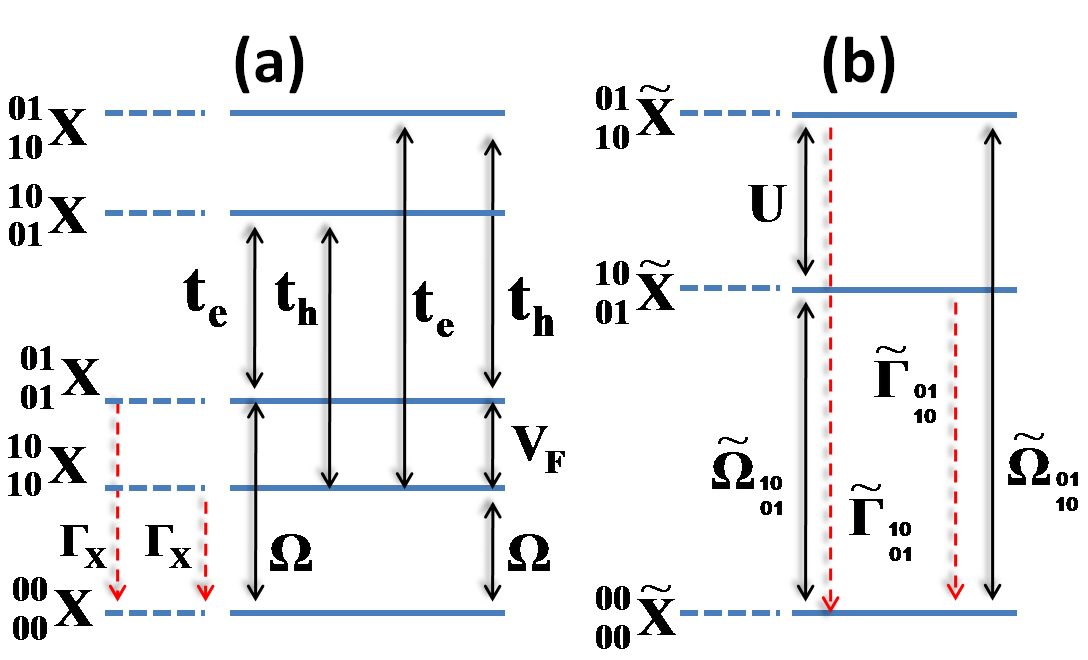}
\caption{\label{levels}(Color online) (a) Bare exciton level configuration corresponding to Hamiltonian Eq.\ (\ref{Ham2}). Solid arrows indicate exciton transitions mediated by different processes of tunneling ($t_e,t_h$), optical pumping ($\Omega$) and F\"{o}rster transfer ($V_F$); dashed arrows represent radiative channels. (b) Effective level configuration according to Eq.\ \ref{Ham3}, after adiabatic elimination of the direct transitions; the tilde on labels represent effective couplings and decay rates.}
\end{figure}

We denote exciton bare states by $\vert _{h_{B}h_{T}}^{e_{B}e_{T}}X\rangle$, where $e_{B(T)},h_{B(T)}=\lbrace 0,1\rbrace$ are the electron and hole occupation numbers, resulting in a total of five states, as shown in Fig.\ \ref{levels}a. The basis of this excitonic Hilbert space contains: the vacuum $|_{00}^{00}X\rangle$; two single {\em direct} exciton states, $|_{10}^{10}X\rangle$ (bottom exciton) and $| _{01}^{01}X\rangle$ (top exciton); and two single {\em spatially indirect} exciton states $|_{01}^{10}X\rangle$ and $|_{10}^{01}X\rangle$. The Hamiltonian in the rotating wave approximation\cite{RWA, Cohen, Shore} is given by
\begin{equation}
{H} = \left(
\begin{array}{ccccc}
\delta_{_{00}^{00}} & \Omega & 0 & 0 & \Omega \\
\Omega& \delta_{_{01}^{01}} & t_{e} & t_{h} & V_{F} \\
0 & t_{e} & \delta_{_{01}^{10}}+\Delta_{S} & 0 & t_{h} \\
0 & t_{h} & 0 & \delta_{_{10}^{01}}-\Delta_{S} & t_{e}\\
\Omega & V_{F} & t_{h} & t_{e} & \delta_{_{10}^{10}}
\end{array}
\right) \, ,\label{Ham2}
\end{equation}
where the columns are associated with the states $|_{00}^{00}X\rangle$,
$|_{01}^{01}X \rangle$, $|_{01}^{10}X \rangle$, $|_{10}^{01}X \rangle$, and
$|_{10}^{10}X \rangle$. Diagonal matrix elements represent detunings of the exciton levels from the laser energy. An applied axial electric field $F$ results in Stark shifts for the indirect exciton detunings $\delta_{_{01(10)}^{10(01)}}=\epsilon_{_{01}^{10}X(_{10}^{01}X)}-\hbar\omega$ given by
\begin{equation}
\Delta_{S}=edF\, .\label{starkshif}
\end{equation}
Likewise, the excitation pulse electric field envelope generates an optical matrix element, $\Omega=\langle _{00}^{00}X\vert \vec{\mu}\cdot\vec{E}\vert _{10(01)}^{10(01)}X\rangle$, for direct exciton transitions.\cite{snoke, foot} Matrix elements, $t_e,t_h$, describe single particle interdot tunnelings for electron and hole respectively.\cite{tunnelings} $V_F$ is an interdot ``hopping" for an entire electron-hole pair, arising from the dipole-dipole interaction coupling two direct exciton states,
\begin{equation}
V_F=\frac{{\mu}_T \, {\mu}_B}{4\pi\epsilon_0\epsilon_{r}d^{3}} \kappa \, ,\label{forster}
\end{equation}
where $\epsilon_{r}$ is the dielectric constant, and $\mu_{T(B)}\sim 6.2 e${\AA} are the interband transition dipole moments. These are assumed parallel to each other (and perpendicular to their separation), which gives an orientation factor $\kappa\sim 1$.\cite{dipoles} In the numerical calculations we use $d\simeq 8.4$nm, which yields a value of $V_F=80\mu$eV. This interaction allows the F\"{o}rster energy transfer mechanism by which a {\em donor} QD transfers its exciton energy to the {\em acceptor} neighboring dot, effectively resulting in the non-radiative interdot ``hoping" of the exciton.\cite{FRET, FRET_Exp_Self, Govorov}

We also consider radiative decay rates, $\Gamma_X^{-1}= \tau_{X} = 1$ns, describing the spontaneous recombination of excitons with spatially direct character.\cite{lifetimes} To that effect, the exciton dynamics is obtained from solutions to the Lindblad master equation, which yields the time evolution for the density matrix of the system,\cite{RWA,Cohen,Shore}
\begin{equation}
\frac{d\rho}{dt} = -\frac{i}{\hbar}[H(t), \rho] + L(\rho)\, .\label{Lindblad}
\end{equation}
The first term on the right describes the coherent evolution of the excitation dynamics, $H(t)$ being the full Hamiltonian of the system,Eq.\ (\ref{Ham2}). The second term, $L(\rho)$, incorporates dissipation processes,
\begin{equation}
L(\rho)=-\sum_j \frac{\Gamma_{ij}}{2}(\lbrace P_j, \rho \rbrace-2\rho_{jj}P_i)\, \label{dissipator}
\end{equation}
where $P_j = \vert j \rangle\langle j \vert$, and $\vert j \rangle$ is an exciton which relaxes into a state $\vert i \rangle$, with rate $\Gamma_{ij}$. The dynamics requires the solution of $N^2$ coupled differential equations for a $N$-dimensional Hilbert space. 

As we will explain in detail later, additional states considered in the model correspond to neutral biexcitonic states, Sec.\ \ref{FRET-Biexciton}, whereas exciton states arising from excited states of the electron (hole) and charged excitons are not considered (assumed to be far removed from the manifold of interest). We also assume that charge tunneling into the contact reservoirs and spin-orbit interactions are negligible, rendering electron-hole exchange decoherence processes unimportant, see Sec.\ \ref{stability}.

\section{Excitonic dressed spectrum}
\label{spectrum}

The interplay of charge tunneling, incident radiation field, and Coulomb interactions, results in coherent interdot coupling. This coupling yields complex molecular states that are superpositions known as the dressed excitonic states. Only a subset of these superpositions leads to allowed transitions, with field dependent amplitudes and energies, which results in anticrossings in the dressed exciton spectra of the QDM. The time evolution of molecular states under optical pumping results in Rabi oscillations, which can be time-integrated to yield the average occupation of the excitons involved in the molecular states. The occupation of such excitons can be probed in principle by differential transmission of a weak probe measuring the population of a particular exciton. It is then possible to construct a level anticrossing (LACS) population map, using the time integrated dynamics of the Hamiltonian, Eq.\ (\ref{Ham2}). For a given map coordinate, $(F,\hbar\omega)$, the integrated population is given by $p_i = (1/t_{L})\int_{0}^{t_{L}}\rho_{ii}(t)dt$, where $t_L\sim 500$ps stands for the constant-amplitude pulse duration, long enough to capture several amplitude oscillations of the exciton populations; only a few Rabi oscillations are necessary to reliably compute $p_i$. Then, any exciton state populated under pumping will exhibit a relative amplitude $p_i(F, \hbar\omega)$ and contribute to features on the corresponding map.
At each coordinate, $(F,\hbar\omega)$, two or more excitons share population if they have non-vanishing components in the dressed state; then by examination of maps corresponding to individual excitons, one can reconstruct the entire dressed spectrum of the system. Alternatively, one can compute the population map of the vacuum state $\vert_{00}^{00}X=|0\rangle$, so that the complete dressed LACS spectrum will correspond to all $(F,\hbar\omega)$ coordinates where this estate is {\em depopulated}; such is the case of Fig.\ 2.
\begin{figure}[htb]
\includegraphics[totalheight=1.0\columnwidth,width=1.0\columnwidth]{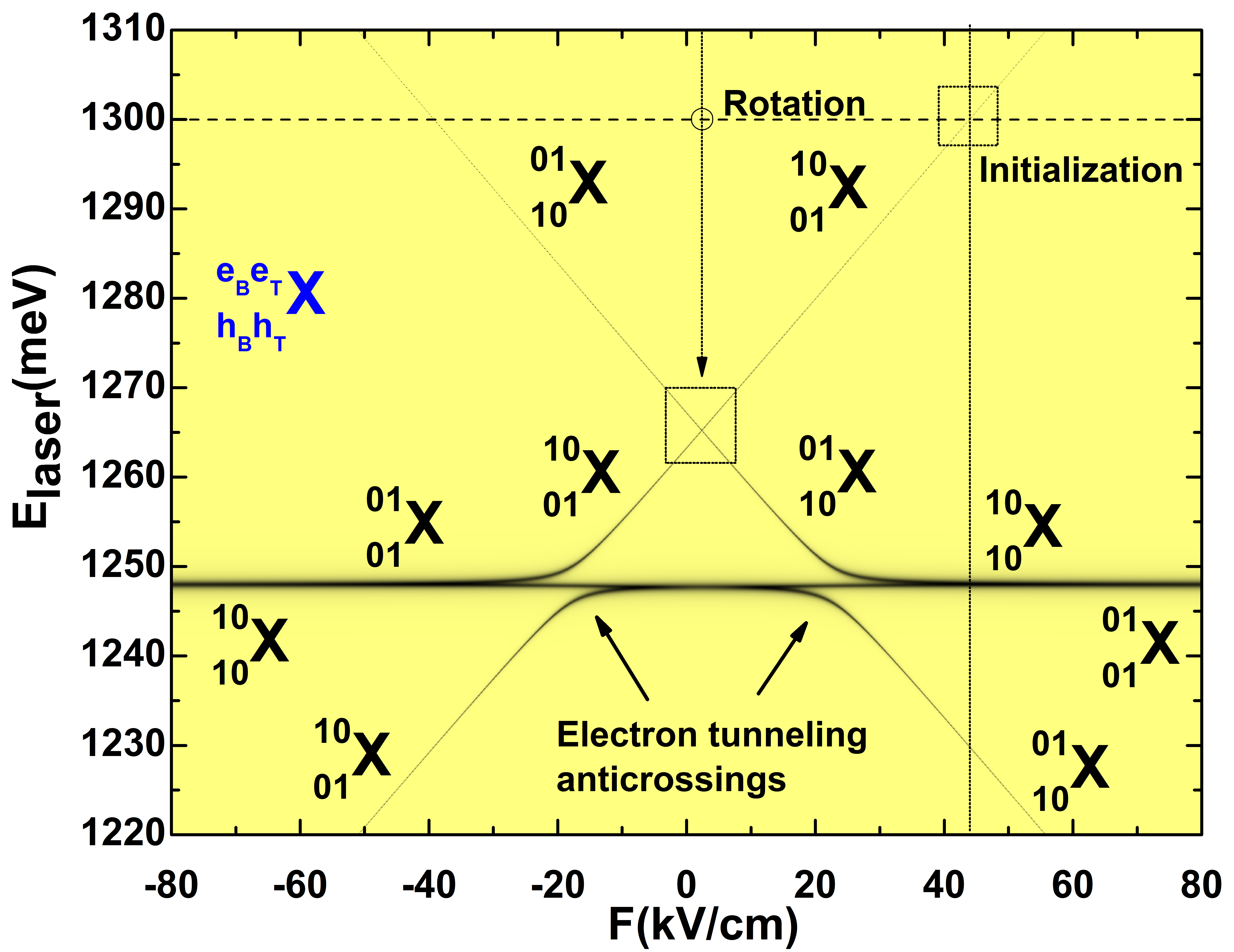}
\caption{\label{populmap}(Color online) Level anticrossing population map of the QDM vacuum state $|_{00}^{00}X\rangle$. As function of applied bias and pump laser energy the vacuum depopulates to other excitons at each resonance. Upper right dashed box indicates resonant excitation into an indirect exciton, $_{10}^{01}X$, at $F_I= 43.4$kV/cm, for a laser energy $E_{laser}=1299.6$meV. The central anticrossing mixes the two indirect excitons at $F_R=2.3$kV/cm. See system parameters in Ref.\ [\onlinecite{numeros}].}
\end{figure}

In contrast, Fig.\ 3a shows the QDM eigenvalue spectrum as function of applied electric field $F$ for a constant value of the pump laser energy, $\hbar\omega = E_{laser}=1299.6$meV and system parameters as in Fig.\ 2. The top right box indicates a level anticrossing that has a correspondent optical signature in the population map of the vacuum state (top right box in Fig.\ 2). Anticrossings at zero energy between the vacuum $|_{00}^{00}X\rangle$ and exciton states $|_{01}^{10}X\rangle$ and $|_{10}^{01}X\rangle$ appear at electric field values, $F \simeq-38.6$ and $43.4$kV/cm, respectively. These values coincide with the corresponding indirect exciton population signatures in upper part of Fig.\ 2, indicated by the dashed horizontal line. In a similar way the central box in Fig.\ 3a encircles a very narrow anticrossing between indirect excitons $|_{01}^{10}X\rangle$ and $|_{10}^{01}X\rangle$, occurring at $F\simeq2.3$kV/cm. Notice that each anticrossing occurs energetically far away from each other, indicating that their eigenstates superpositions are only weakly coupled to the others. This suggest that an effective Hamiltonian, represented in the basis of the qubit subspace, should reproduce these anticrossing signatures.


\section{Qubit Extraction}
\label{extraction}

In the spectroscopy of QDMs, the appearance of a level anticrossing signature points to the onset of an important interdot interaction. These interactions are in some cases not straightforwardly explained by the off diagonal matrix elements of the Hamiltonian, $H_{ij}$, connecting two allowed states. For example, two excitons $|i\rangle$ and $|j\rangle$, with a very weak oscillator strength can couple strongly to the radiation field via higher order transitions mediated by non-optical processes, such as charge tunneling. Such is the case for indirect excitons in our model, which couple via higher order processes to the radiation field, yet they are assumed to have zero oscillator strength. These states have well-defined optical signatures in the LACS map in Fig.\ 2 and exhibit an anticrossing with the vacuum in the eigenvalue spectrum in Figs.\ 3b and 3d.

This coupling of two indirect excitons via higher order transitions, and their respective coupling to the vacuum, should be revealed by an effective Hamiltonian that describes the same physics as the original, but constrained to a sector of the Hilbert space whose wave functions correspond just to the eigenvalue spectrum in the anticrossing region. This projected Hamiltonian is of reduced dimensionality and should have non-zero off-diagonal matrix elements connecting the states involved. We employ the Feshbach projection operator formalism,\cite{Feshbach} which permits the derivation of an effective Hamiltonian with exciton eigenstates of pure indirect character. The direct exciton sector of the Hamiltonian is adiabatically eliminated (``projected out"), and its dynamical effects become embedded in the matrix elements of the effective Hamiltonian. The requirements of adiabatic elimination are satisfied by two conditions: (1) adiabatic variation of all external fields (2) ability to isolate spectrally, by tuning the excitation energy (off-resonant condition) and applied electric field, the confluence of two excitons from the remaining exciton manifolds. The Hamiltonian obtained by projection, would describe the time evolution of qubits in the unitary and dissipative regimes.

In the following discussion, for simplicity we consider a \textit{closed} quantum system with the Hamiltonian given by Eq.\ \ref{Ham2}. The Hamiltonian can be separated in to two parts, $H=H_{0}+V$, where $H_{0}$ is the unperturbed diagonal part, and $V$ a perturbation. Let $\mathcal{P}$ be the relevant subspace spanned by the excitons that self-avoid at a chosen system resonance. In the same way let $P$ and $Q=1-P$ be projector operators onto and outside of $\mathcal{P}$, respectively. The effective Hamiltonian is given by \cite{Feshbach}
\begin{equation}
\tilde{H}(z)=PH_{0}P+PR(z)P \, ,
\end{equation}
with $z=E \pm i\epsilon$, where $E$ and $\epsilon$ are the real and imaginary parts of the complex energy eigenvalue $z$. The first term of $\tilde{H}$ is the leading unperturbed part of the Hamiltonian inside $\mathcal{P}$, with the second term containing the level shift operator, $R(z)=V+VQ[z-QH_{0}Q-QVQ]^{-1}V$, projected onto $\mathcal{P}$. The latter term can be seen as a Hamiltonian that permits the calculation of the energy level shifts with respect to the unperturbed levels. Allowing the Hamiltonian to depend on its eigenvalues $z$, makes the eigenvalue equation non-linear. Additionally, analytic continuation of the eigenvalues into the complex plane allows the introduction of non-Hermitian Hamiltonians that incorporate dissipation processes taking place outside the relevant subspace, $\mathcal{P}$. Self consistent solutions to the non-linear eigenvalue equation are used to obtain the eigenvalue spectrum in the vicinity of a level crossing and anticrossing. Near a level anticrossing (and in the absence of accidental degeneracies) there is a unique self-consisting solution of $z(F)$ for each value of the applied electric field $F$.

\begin{figure}[htb]
\includegraphics[totalheight=1.0\columnwidth,width=1.0\columnwidth]{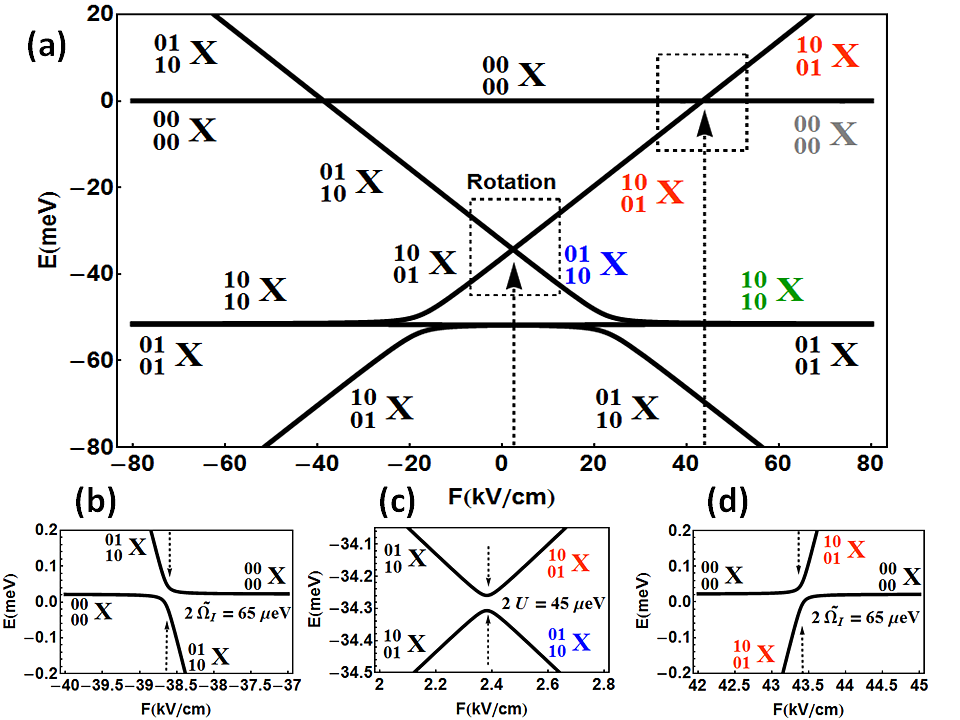}
\caption{\label{eigenvalues}(Color online) (a) Dressed exciton eigenvalue spectrum at fixed excitation energy $\hbar\omega=1299.6$meV. Boxes show anticrossings associated to couplings to the radiation field and between indirect excitons. These features have a one to one correspondence to the optical signatures in Fig.\ \ref{populmap}. (b) and (d) zoom into anticrossings near $E=0$. (c) is the central anticrossing that mixes the indirect excitons spanning the qubit subspace.}
\end{figure}

\subsection{Indirect exciton qubit Hamiltonians}
\label{3levels}

In order to obtain the dynamics of molecular excitons with spatially indirect character, we project the total Hamiltonian (\ref{Ham2}) onto the three level system, $|_{00}^{00}X\rangle,|_{01}^{10}X\rangle,|_{10}^{01}X\rangle$, shown in Fig.\ 1b. The resulting Hamiltonian incorporates an effective coupling among the indirect excitons, $U$, and effective couplings of indirect excitons to the radiation field, $\tilde{\Omega}_{_{10}^{01}}$ and $\tilde{\Omega}_{_{01}^{10}}$, respectively. We assume that the direct exciton levels are resonantly coupled by FRET, $\delta_{_{10}^{10}}=\delta_{_{01}^{01}}=\Delta$, and that the bottom and top QDs couple with the same strength to the radiation field, $\Omega$. This assumption yields,  $\tilde{\Omega}_{_{10}^{01}} = \tilde{\Omega}_{_{01}^{10}}= \tilde{\Omega}_I$. Let $\mathcal{P}_{\Lambda}$ be the subspace subtended by the vacuum and the indirect exciton levels; then a projection of Hamiltonian (\ref{Ham2}) onto $\mathcal{P}_{\Lambda}$ gives
\begin{equation}
\tilde{H}^{(\Lambda)}(z)=\left(
                       \begin{array}{ccc}
                         \Delta_{_{00}^{00}}(z) & \tilde{\Omega}_I(z) &  \tilde{\Omega}_I(z) \\
                         \tilde{\Omega}_I(z) & \Delta_{_{01}^{10}}(z)+\Delta_S & U(z) \\
                         \tilde{\Omega}_I(z) & U(z) & \Delta_{_{10}^{01}}(z)-\Delta_S \\
                       \end{array}
                     \right)\, .\label{Ham3}
\end{equation}
The matrix elements of $\tilde{H}^{\Lambda}(z)$ contain level shift detunings  $\Delta_i$ and effective couplings $U$ and $\tilde{\Omega}_I$. The indirect exciton effective detunings are given by
\begin{equation}
\Delta_{_{01}^{10}(_{10}^{01})}(z)=\delta_{_{01}^{10}(_{10}^{01})}+\delta_I(z)\, ,\label{indshift-1}
\end{equation}
\begin{equation}
\delta_I(z)=\frac{(z-\Delta)(t_e^2+t_h^2)+2t_et_hV_F}{(z-\Delta)^2-V_F^2}\, ,\label{indshift}
\end{equation}
while the shift of the zero of energy (we have set $\delta_{00}^{00}=0$) is given by
\begin{equation}
\Delta_{_{00}^{00}}(z)=-\frac{2\Omega^2}{\Delta+V_F-z}\, .\label{zeroshift}
\end{equation}
correspondingly, the indirect exciton effective coupling to the radiation field are both given by
\begin{equation}
\tilde{\Omega}_I(z)=-\frac{\Omega(t_e+t_h)}{\Delta+V_F-z}\, .\label{radcoupling}
\end{equation}
Figures 3b and 3d show the anticrossing gaps opened by the couplings $\tilde{\Omega}_I$ in Eq.\ (\ref{Ham3}). Both gaps have a width $2\tilde{\Omega}_I=65 \mu$eV, occurring at applied electric field values of $F\simeq -38.6$ and $43.4$kV/cm, respectively.\cite{numeros} $\tilde{\Omega}_I$ is directly proportional to the direct transition dipole matrix element $\Omega$ and tunneling amplitudes $(t_e+t_h)$. In other words, molecular indirect excitons are provided with an effective oscillator strength when the dots in the QDM are tunnel coupled. This leads to the possibility of resonant excitation of the indirect excitons, even if they have a vanishing intrinsic oscillator strength. This explains the ``lighting up" of indirect excitons in Fig.\ 2 when optically driving the QDM at electric field values $\vert F\vert\geq 20$kV/cm.

The relevant interaction between two neutral indirect excitons is given by
\begin{equation}
U(z)=\frac{2(z-\Delta)t_{e}t_h+(t_{e}^2 + t_{h}^2) V_F}{(z-\Delta)^2 - V_F^2}\, ,\label{intracoupling}
\end{equation}
which dominates for values of electric field $F_R=2.3$kV/cm. Eq.\ (\ref{intracoupling}) shows that $U$ is \textit{independent} of the laser intensity embodied in the direct dipole matrix element $\Omega$. This represents an important desirable feature for a qubit defined in the indirect exciton subspace. It implies that for a fixed laser energy, the effective qubit subspace generated by the indirect excitons $|_{01}^{10}X\rangle$ and $|_{10}^{01}X\rangle$ is effectively shielded against the external disturbance of the intense optical field and less susceptible to the effects of spontaneous direct exciton recombination. On the other hand, $U$ arises predominantly from electron and hole tunneling, with a weak contribution from FRET; this means that the molecular indirect subspace evolves mainly by its internal dynamics. For the system under consideration, see Ref.\ [\onlinecite{numeros}], we find $2U\simeq 45\mu$eV.

A more concise qubit Hamiltonian is obtained by projecting Eq.\ (\ref{Ham2}) onto $\mathcal{P}_I=\lbrace|_{01}^{10}X\rangle,|_{10}^{01}X\rangle\rbrace$. Then, one obtains a two-level Hamiltonian describing the spectrum at the central anticrossing when $U$ dominates. The projection results in
\begin{equation}
\tilde{H}^{(I)}(z) = \left(
                     \begin{array}{cc}
                       H_{22}^{\Lambda}(z) & U(z) \\
                       U(z) & H_{33}^{\Lambda}(z) \\
                     \end{array}
                   \right)+\xi(z)(\sigma_X+I)\, ,\label{Ham4}
\end{equation}
in terms of the matrix elements of Eq.\ (\ref{Ham3}), and corrections with $\sigma_X$ and $I$ being the x-Pauli and identity matrices, respectively. The correction term $\xi(z)=\frac{\tilde{\Omega}_I^{2}(z)}{z-\Delta_{_{00}^{00}}(z)}$ describes the small perturbations arising from optical excitation and reflects the fact that the effective subspace is not perfectly isolated when the radiation field is on. Notice, however that the correction terms disappear in the absence of pumping ($\Omega=\tilde{\Omega}_I=\xi=0$).  We also emphasize that $\tilde{H}^{(I)}$, defined on the subspace $\mathcal{P}_I$, provides a better description of the qubit rotation, while $\tilde{H}^{(\Lambda)}$ is more adequate and convenient  for describing the initialization of the qubit system (via the effective coupling of indirect excitons to the light field).  However, both subspaces and associated Hamiltonians are suitable to describe the intrinsic qubit dynamics.

\subsection{FRET and biexciton effects}
\label{FRET-Biexciton}
As mentioned above, when direct exciton transitions in the two dots are near resonant, the F\"{o}rster energy transfer mechanism plays an important role on the direct exciton superradiant dynamics.\cite{Al-Ahmadi,Sitek} Typically $V_F\simeq 0.08$meV, for interdot separation $d\simeq 8.4$nm. This is a small value in comparison with electron tunneling $t_e$,\cite{tunnelings} but appreciable enough to split the direct exciton spectral lines and redistribute the exciton population (spectral weight) among the molecular states in a steady state regime.\cite{Rolon}
Interestingly, the denominators in the effective Hamiltonian matrix elements in Eq.\ (\ref{radcoupling}) and (\ref{intracoupling}) exhibit a dependence on $V_F$ as a correction to the direct exciton detuning $\Delta$. Therefore, any influence of $V_F$ can be strongly suppressed whenever $\Delta\gg V_F$. In this regime, possible dephasing effects due to FRET would be suppressed as well. In our model, $\vert\Delta|\simeq 51.5$ meV, assuring the indirect exciton qubit subspace is indeed shielded against the perturbation effects of FRET.

On the other hand, strong laser excitation can pump additional exciton levels outside the relevant subspace of consideration. The closest excitations are biexciton resonances, which cannot be in principle ignored in the dynamics of single excitons, as their detuning is at most a few meV.\cite{Biexciton} The pumping of biexcitons in either QD, $\vert_{20}^{20}X\rangle$, $\vert_{02}^{02}X\rangle$, expands the bare exciton basis to 14 states $\vert _{h_{B}h_{T}}^{e_{B}e_{T}}X\rangle$, with possible double occupancy of the single particle levels, $e_{B(T)},h_{B(T)}=\lbrace 0,1,2\rbrace$, which becomes more significant with higher excitation power and/or short laser pulses. However, the detuning of the biexciton levels, and the need for a direct exciton prior to its formation, result in weak perturbative effects of the biexciton level manifolds for the values of the matrix element $\Omega$ considered here. Moreover, the biexciton manifolds decouple once the excitation power switches off during the dynamical control procedure, see Sec.\ \ref{control}. Other excitations, such as LO phonon resonances, appear $\sim 35$meV above the lowest exciton transition for GaAs, and can be safely ignored.\cite{LOPhonon} 
We notice that the chosen structure parameters (QD confinement sizes) result in excited electron and hole states (and associated excitons) far from the relevant anticrossing gaps, $2U$ and $2\Omega_I$, so that these other excitations can be safely ignored (see Sec.\ \ref{stability}). 

\section{Coherent rotation of Indirect exciton Qubit}
\label{control}

In what follows, we consider the system as a fully \textit{open} quantum system and consider explicitly radiative recombination of direct excitons. In this sense, we analyze our results in terms of numerical solutions to the Lindblad master equation, Eq.\ \ref{Lindblad}, with all 14 excitonic states included. Our discussion of the projected subspace $\mathcal{P}_I$, indicates that the molecular indirect exciton subspace is indeed weakly influenced by interdot energy transfer mechanisms, $V_F$, and excitation power, $\Omega$. This suggests that we can achieve control of the indirect exciton qubits by tuning the effective coupling strengths $\tilde{\Omega}_I$, $U$ and by application of external time dependent electric fields (we will discuss in detail dissipation processes in Sec.\ \ref{dissipation} below).\cite{Destefani} We use a cyclic adiabatic variation of the applied field, $F(t)$, at fixed excitation energy $\hbar\omega $, between a regime where the system effectively contains two-levels mixed by the coupling $\tilde{\Omega}_I$, into a regime where the system contains two levels mixed by $U$. One can use short adiabatic bias pulses for qubit initialization and rotation operations.\cite{Gorman-Yamamoto} Figure 4a shows a cyclic sweep of applied bias, the left arm ($0 \leq t \leq 0.17$ns) indicates the initialization regime, shown in more detail in Fig.\ 4c. The slow forward bias ramp ($0.17 \leq t \leq 1.22$ns) drives the system into the qubit rotation regime, see Fig.\ 4d. The plateau in the bias pulse ($1.22 \leq t \leq 2.99$ns) corresponds to the rotation regime, and its tunable duration determines at which particular time one decides to rotate the input state (red curves) or not. If rotated, the reverse bias adiabatic ramp ($ 2.99 \leq t \leq 4.05$ns) transfers the output state population (blue curves) into a direct exciton (green curves), which in turn depopulates subsequently into the vacuum (dashed curves) for $(t \geq 4.05)$ns, see Fig.\ 4e. The overall dynamics of the implemented coherent control is shown in Fig.\ 4b. In what follows, let us discuss each region in more detail. 

\subsection{Initialization}
\label{initialization}

The couplings $U$ and $\tilde{\Omega}_I$ dominate in different field regimes, therefore the representation subspace of Hamiltonian Eq.\ (\ref{Ham3}) can be decoupled in three different regions. When $\tilde{\Omega}_I$ dominates, for large values of electric field and positive energy detuning, we can construct two projected subspaces, spanned by the basis vectors $\lbrace|_{00}^{00}\rangle, |_{01}^{10}X\rangle\rbrace$ and $\lbrace|_{00}^{00}\rangle, |_{10}^{01}X\rangle\rbrace$, respectively. For an excitation energy of $\hbar\omega=1299.6$meV, coherent Rabi oscillations are induced in each of these subspaces for applied electric field values of $F=-38.6$ and $43.4$kV/cm, which corresponds to resonant excitation at either level anticrossing, shown in figures 3b and 3d, respectively. This allows the implementation of \textit{$\pi$ rotations} within a time interval $\frac{\pi}{2\tilde{\Omega}_I}\simeq 63.6$ps, enabling the possibility of initializing the system in either of the logical states, $|_{10}^{01}X\rangle$ or $|_{10}^{01}X\rangle$, respectively. On the other hand,  $\Omega_I$ depends inversely on the direct exciton detuning, $\Delta$, which sets an upper bound such that $\Omega_I\gg\Gamma_X$, since otherwise the fidelity would be hampered by spontaneous recombination.

Figure 4c shows the initialization of the indirect exciton $|_{10}^{01}X\rangle$ by a $3\pi$ rotation. When driving the system at the anticrossing, corresponding to the coordinate ($F_I=43.4$kV/cm, $\hbar\omega=1299.6$meV), the initialization takes place after switching off the pulsed resonant excitation at a time $t = 200$ps. The initialization occurs with near unity fidelity, $F=\langle_{01}^{10}X|\rho(t_i)|_{01}^{10}X\rangle\simeq0.97$, due to the almost perfect isolation of the subspace $\mathcal{P}_{\Lambda}$.

\begin{figure}[htb]
\includegraphics[totalheight=1.0\columnwidth,width=1.0\columnwidth]{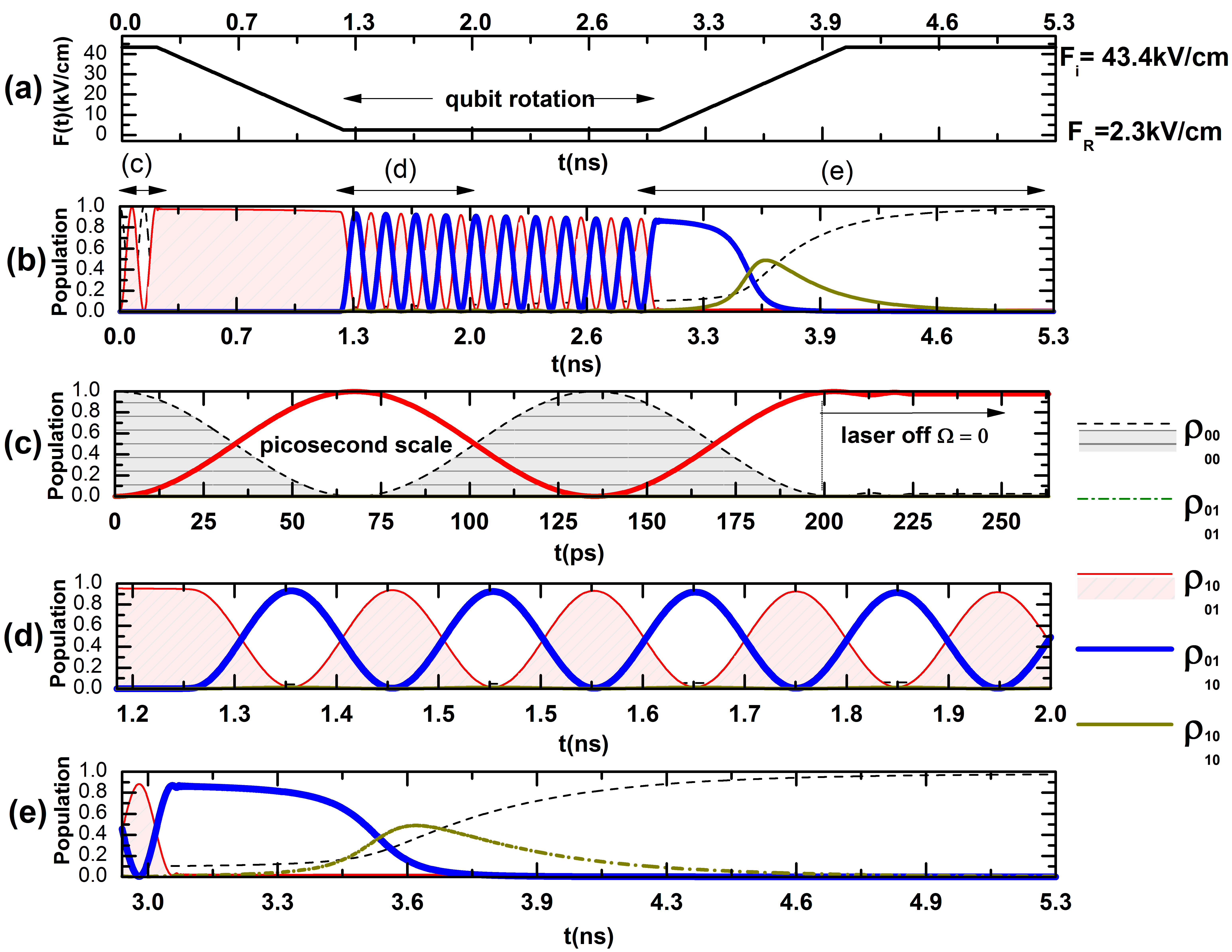}
\caption{\label{dynamics}(Color online) Exciton population dynamics subject to an applied cyclic bias pulse. (a) Applied bias pulse sweep of duration $\tau\simeq 5.5$ns, the cycle sweeps the interval $F=[43.4,2.3]$kV/cm. (b) Exciton population dynamics corresponding to the cyclic sweep above. Notice (a) and (b) panels have the same time scale, while (c)-(e) show details of different regions, as indicated. (c) Rabi oscillations between $\vert _{00}^{00}X\rangle$ and $\vert _{01}^{10}X\rangle$ excitons. After a $3\pi$ rotation, at $t= 200$ps, the laser is turned off and qubit is initialized with near unity fidelity. (d) Rabi oscillations inside the qubit subspace. (e) Read out scheme of the output qubit state, via tunneling adiabatic passage into a direct exciton. Depopulation of the output state occurs into the vacuum state.}
\end{figure}

\subsection{Qubit rotation}
\label{rotation}

In order to perform a desired rotation operation involving input states $|_{01}^{10}X\rangle$ and $|_{10}^{01}X\rangle$, the system is adiabatically driven into the central anticrossing occurring at $F_R\simeq 2.3$kV/cm, see central box region in Fig.\ 3a. Figure \ref{dynamics}a shows a sweep that drives the system from the anticrossing at $F_I\simeq 43.4$kV/cm to the one at $F_R$. There, the system evolves by its internal dynamics. This is emphasized by the absence of the coupling $\Omega$ in $U$, and the switch off of the laser once the qubit is initialized. The coherent oscillations allow for qubit rotations in the picosecond scale,\cite{Ramsay} with a characteristic time $\frac{\pi}{2U}=91.8$ps, and exhibit a near unitary amplitude within the time frame shown in Fig.\ 4d. Notice that the coherent oscillation relaxes on a longer time scale. This is due to strong direct exciton relaxation rates, and the weak effects of exciton virtual transitions occurring outside the qubit subspace, on the matrix element $U(z)$ and energy shift of the indirect exciton, $\delta_I(z)$, see Eqs.\ (\ref{intracoupling}) and (\ref{indshift}), respectively. The duration of this rotation determines how much population is transferred into the output state, in other words, how much the final indirect molecular state would follow the eigenvalue line, $\vert_{10}^{01}X\rangle$, upon bias reversal.

\subsection{Readout}
\label{readout}
Once the qubit rotation has taken place, and the population is transferred into the output state after a $\pi$ rotation, the molecular eigenstate follows a different running eigenvalue in reverse bias. This is observed in Fig.\ 4e after $t=3$ns; here the applied bias pulse drives the output state $|_{10}^{01}X\rangle$ along the dressed spectral line $\vert_{10}^{01}X\rangle$, starting at $F_I\simeq 2.3$ and finishing at $F_R\simeq 43.4$kV/cm. The fidelity of the readout depends on a conditional adiabatic population passage\cite{adiabatic, conditional} from $\vert_{01}^{10}X\rangle$ into the direct exciton $|_{10}^{10}X\rangle$, then with the partial population transfer into the vacuum $|_{00}^{00}X\rangle$ (green solid and black dashed line, Fig.\ 4e). At the end of the sweep, far away from the central anticrossing, the direct exciton $|_{10}^{10}X\rangle$ is depopulated, by recombination emitting luminescence, without perturbing the indirect states or any other nearby exciton.

\subsection{Stability of coherent control}
\label{stability}

We should emphasize that the control scheme requires the central indirect-exciton anticrossing to be isolated from other exciton states and resonances associated with transitions out of the qubit subspace. For a wide range of system parameters, the coherent rotation regime is achieved via an anticrossing which appears isolated in an energetically narrow window (easily detunable from other transitions), and occurs even if the ground states of the QDM dots are non-resonant ($\delta_{10}^{10}\neq\delta_{01}^{01}$), protecting the qubit subspace and enhancing coherence. If other excited states (such as those associated with excited electron/hole levels of the molecule) appear in the vicinity of the qubit window, their effects can be naturally incorporated in the description.  They may result in changes of the initialization field and pumping, but they would not intrinsically affect the main qubit rotation scheme. 
Certainly, strong distortion of the relevant anticrossings, Figs.\ 3c-d, by a nearby state affects the rotation and initialization fidelity, since in that case the system would not be approximated by a well-separated two-level system. Further complications could arise if charge tunneling rates into the diode contacts compete with the control time scales; in that case, charged excitons (negative and positive trions) would not be negligible, affecting the charge stability of the exciton qubits, and enabling decoherence by exciton spin dephasing mediated by electron-hole exchange interaction.\cite{Falt} However, these constraints can be relaxed by proper geometrical engineering of the excitonic spectrum and excitation conditions, and by selection of suitable molecules among the many produced in typical processes.  

\section{Dissipation Effects}
\label{dissipation}

We have assumed that the neutral indirect excitons are optically inactive for the chosen interdot distance.\cite{numeros} So that their intrinsic recombination rate is $\Gamma_I = 0$. However, we have found that even with this assumption, the effect of the direct exciton spontaneous recombination, plus the influence of virtual transitions (mediated by tunneling) occurring outside the qubit subspace, provides the molecular indirect exciton with a \textit{finite} effective oscillator strength and lifetime, which is ultimately a consequence of interdot quantum coupling.

\begin{figure}[htb]
\includegraphics[totalheight=0.80\columnwidth,width=1.0\columnwidth]{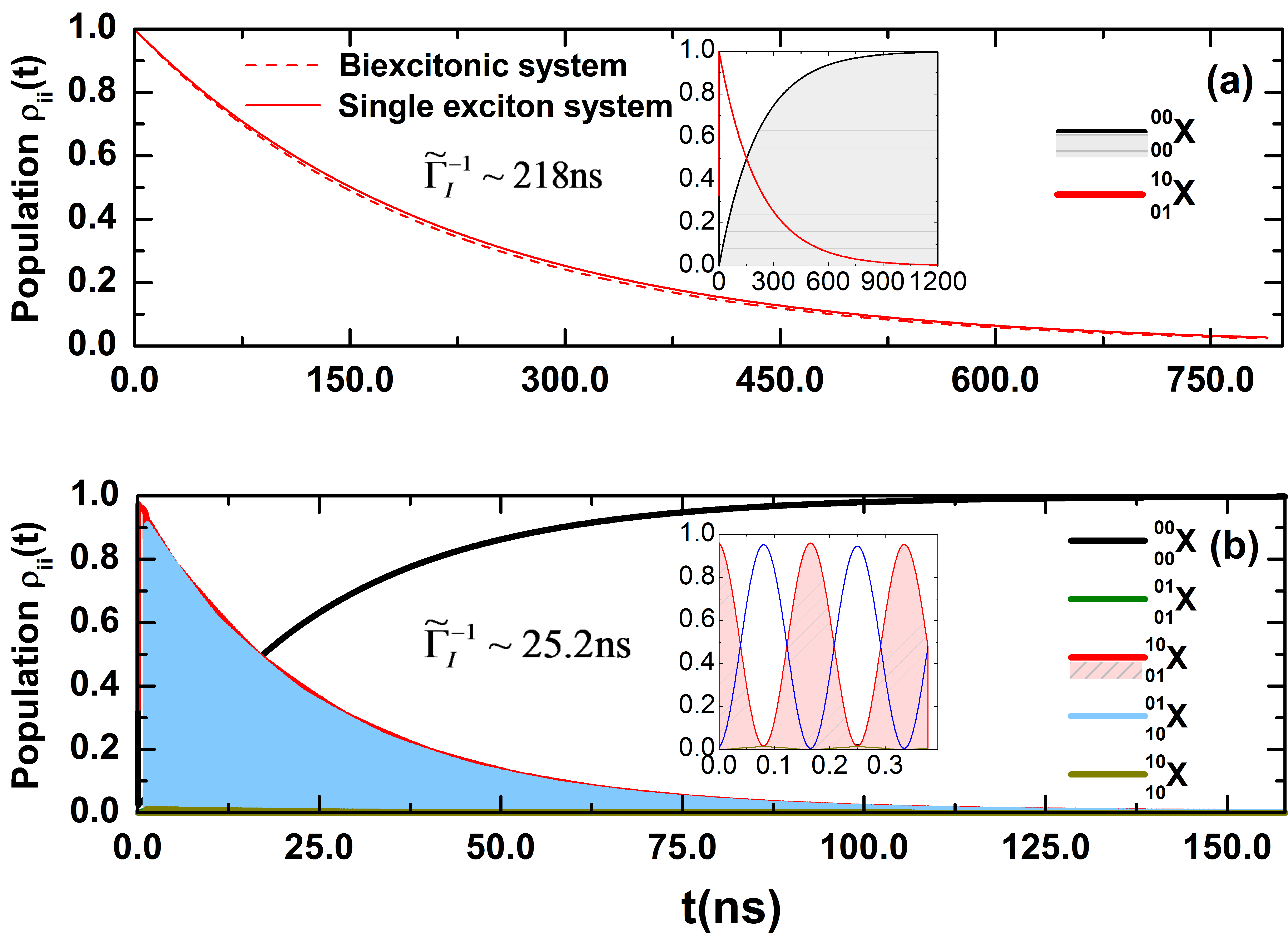}
\caption{\label{Relaxation}(Color online) Population relaxation dynamics for indirect excitons, from the numerical solutions of Eq.\ 4 with up to 14 exciton states included. (a) Population decay after switching off the laser light in the initialization regime at $F_I=43.4$kV/cm. Dashed line indicates the case where biexciton states are taken into account. Inset shows the corresponding depopulation into the vacuum. (b) QDM internal dynamics in absence of optical perturbations, shows decay of Rabi oscillations of $\vert_{01}^{10}X\rangle$ and $\vert_{10}^{01}X\rangle$ (red line, blue line) in the qubit rotation regime at $F_R=2.3$kV/cm. Inset shows Rabi flops in the early stage of the dynamics}
\end{figure}

Figure 5a shows the population time dependence of the input state $\vert _{10}^{01}X\rangle$ after switching off the excitation power, $\Omega(\tau)=0$, at a time $\tau=200$ps, as in Fig.\ 4c (from the numerical solution of Eq.\ 4, with up to 14 exciton states included). We see that the population relaxes into the vacuum (see inset) with a lifetime $\tilde{\Gamma}_{I}^{-1}\simeq 218$ns. Comparatively the dashed line indicates depopulation of the state when the biexcitonic degrees of freedom (and all 14 exciton basis states) are taken into account (notice they have no influence on the relaxation time when  $\Omega=0$).
Figure 5b shows the population time dependence of the QDM, when the input state has been driven adiabatically into the rotation regime. Here the system is driven solely by its internal dynamics, which arises due to strong electron and hole tunneling, and the small influence of FRET. Both indirect excitons (blue and red solid lines) enter a coherent oscillation regime, relaxing into the vacuum with a lifetime $\tilde{\Gamma}_{I}^{-1}\simeq 25.2$ns, with vanishing population transfer into the direct excitons. Notice that the relevant coherent oscillation period ($\simeq 91.8$ps), is orders of magnitude shorter than the relaxation time. On the other hand, our results indicate that despite assuming infinite lifetime for the bare indirect excitons at very large bias, $\vert F\vert\gg1$kV/cm, the interdot couplings provide a strong dissipation channel for indirect excitons to relax into the vacuum at any finite value of $F$. In fact, experimental results have shown electric field tuning of radiative lifetimes at the direct-indirect molecular exciton resonance (near the tunneling induced anticrossing, $F\simeq \pm 20$kV/cm), still in the range of $\sim2$ns to $\sim10$ns, if acoustic phonon mediated inter-level relaxation processes are present.\cite{Nakaoka}

Calculation of the population damping of the state $\vert_{01}^{10}X\rangle$ inside the effective subspace $\mathcal{P}_{\Lambda }$, as well as the damping of coherent oscillations within the qubit subspace $\mathcal{P}_I$, gives the effective lifetime of the qubit input state during initialization and the decoherence time during qubit rotation, respectively. Although we consider just the effects of intrinsic spontaneous recombination of the direct excitons, the effective decay rates of the molecular states can be significatively different, since they depend strongly on all interdot coupling mechanisms, and thus can be tuned at will as function of laser detuning, pump power and applied electric field.

In what follows we are interested in the dissipative dynamics of the indirect excitonic subspace. Since $V_F\ll\vert\Delta\vert$, we can ignore the effects of FRET in the following discussion. In order to obtain an analytical expression for the effective decay rates, we start with the non-Hermitian Hamiltonian\cite{Cohen, Shore}
\begin{equation}
H_\Gamma= H-\frac{i\Gamma_X}{2} (\vert _{10}^{10}X\rangle\langle _{10}^{10}X \vert + \vert _{01}^{01}X\rangle\langle _{01}^{01}X \vert )\, ,\label{non-Hermitian}
\end{equation}
and project it onto the subspaces $\mathcal{P}_{\Lambda}$ and $\mathcal{P}_I$.
Consequently the resultant effective Hamiltonian, $\tilde{H}_{\Gamma}^{(I)}(z)$, will be non-Hermitian, with matrix elements that depend on the direct exciton decay, $\Gamma_X$. Projection onto the subspace $\mathcal{P}_I$, results in
\begin{equation}
\tilde{H}_{\Gamma}^{(I)}(z)=\tilde{H}_{0}^{(I)} + \tilde{H}_{Re}^{(I)}(z)+i\tilde{H}_{Im}^{(I)}(z)\, ,\label{HPInonH}
\end{equation}
which must of course reduce to Eq.\ (\ref{Ham4}) for $\Gamma_X=V_F=0$. The diagonal part of the non-Hermitian projected Hamiltonian is,
\begin{equation}
\tilde{H}_{0}^{(I)}(z) = \left(
                     \begin{array}{cc}
                       \delta_{_{01}^{10}}+\Delta_S  & 0 \\
                       0 & \delta_{_{01}^{10}}-\Delta_S
                     \end{array}
                   \right)\, ,\label{H0nonH}
\end{equation}
with a projected real perturbation given by
\begin{equation}
\tilde{H}_{Re}^{(I)}(z) =\frac{1}{\Gamma_X^2+4(z-\Delta)^2} \left(
                     \begin{array}{cc}
                       \alpha(z)  & \upsilon(z) \\
                       \upsilon(z) & \alpha(z)
                     \end{array}
                   \right)\, ,\label{HRenonH}
\end{equation}
and its corresponding  imaginary perturbation given by
\begin{equation}
\tilde{H}_{Im}^{(I)}(z) =\frac{1}{\Gamma_X^2+4(z-\Delta)^2} \left(
                     \begin{array}{cc}
                       \beta  & \gamma \\
                       \gamma & \beta
                     \end{array}
                   \right)\, ,\label{HImnonH}
\end{equation}
with matrix elements given by,
\begin{equation}
\begin{array}{cc}
\alpha(z) = 4(t_e^2+t_h^2)(z-\Delta), & \beta = -2(t_e^2+t_h^2)\Gamma_X \\
\\
\upsilon(z) = 8t_et_h(z-\Delta), & \gamma = -4t_et_h\Gamma_X\, .\label{NHH-elements}
\end{array}
\end{equation}

\begin{figure}[htb]
\includegraphics[totalheight=0.65\columnwidth,width=1.0\columnwidth]{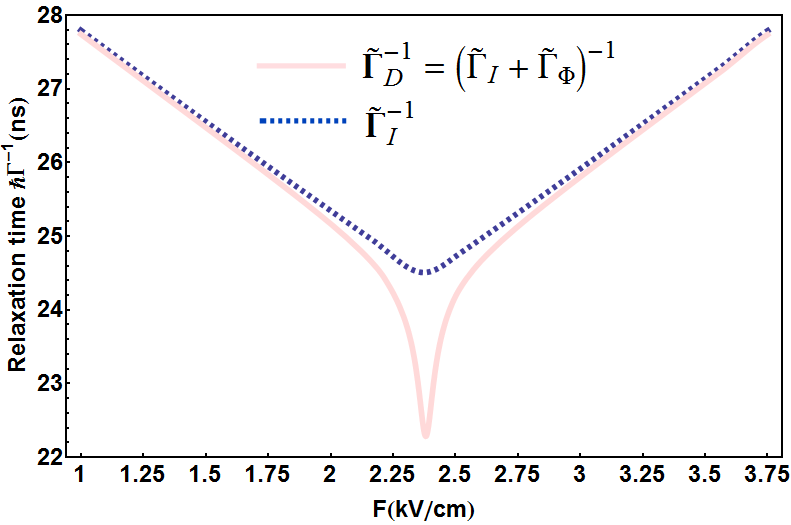}
\caption{\label{EffectiveGammas}(Color online) Relaxation times for bare and dressed indirect excitons as function of applied electric field $F$. Light red (blue) line corresponds to relaxation time for the dressed (bare) indirect exciton eigenstates. Far from anticrossing value of electric field $F_I=2.38$kV/cm, the dressed relaxation time approaches the bare exciton relaxation time as molecular indirect eigenstates of $\tilde{H}_{\Gamma}^{(I)}$ become the indirect states $\vert_{10}^{01}X\rangle$, $\vert_{01}^{10}X\rangle$.}
\end{figure}

The dissipative part of the Hamiltonian, $\tilde{H}_{Im}^{(I)}$, contains two sources of decoherence, a diagonal source proportional to $\beta$ and an off-diagonal source proportional to $\gamma$. $\beta$ is the intrinsic population relaxation rate of the dressed indirect excitons, which is non-vanishing when either the electron or the hole tunneling is non-zero. This term is the dominant part of the relaxation rate, and is present along the whole electric field sweep, inside and outside the qubit subspace. $\gamma$ arises from the simultaneous tunneling of an electron and a hole, and can be interpreted as an interference term between the simultaneous paths of these two charges. This term is much smaller (as $t_e \gg t_h$), and contributes mostly inside the qubit subspace. Therefore, the damping of coherent oscillations during the rotation operation, contains population decay and dephasing.

The decay rates of the dressed states inside the qubit subspace, $\mathcal{P_I}$, can be obtained by diagonalization of the Hamiltonian (\ref{HPInonH}), which gives the total imaginary component of the indirect dressed excitons. This total rate can be written as $\tilde{\Gamma}_D(z)=$ Im[Diag$( \tilde{H}_{\Gamma}^{(I)})$]$_{jj}$ $=\tilde{\Gamma}_I(z) + \tilde{\Gamma}_{\Phi}(z)$, for either indirect exciton $\vert j \rangle$, where

\begin{equation}
\tilde{\Gamma}_I(z) = \frac{2(t_e^2+t_h^2)\Gamma_X}{\Gamma_X^2+4(z-\Delta)^2} \, ,
\end{equation}
\begin{widetext}
\begin{equation}
\tilde{\Gamma}_\Phi(z)=\left(\left(\frac{\delta_{_{01}^{10}}-\delta_{_{10}^{01}}}{2}+\Delta_S\right)^4 +\frac{8t_e^2t_h^2(32t_e^2t_h^2-(\delta_{_{01}^{10}}-\delta_{_{10}^{01}}+2\Delta_S)^2(\Gamma_X^2-4(z-\Delta)^2))}{(\Gamma_X^2+4(z-\Delta)^2)^2}\right)^{\frac{1}{4}}\sin \theta(z)\, ,
\end{equation}
\begin{equation}
\theta(z)=\frac{1}{2}\arctan \left(\left(\frac{16t_et_h}{\delta_{_{01}^{10}}-\delta_{_{10}^{01}}+2\Delta_S}\right)^2 \frac{(\Delta-z)\Gamma_X}{(\Gamma_X^2+4(z-\Delta)^2)^2 -\left( \frac{8t_et_h}{\delta_{_{01}^{10}}-\delta_{_{10}^{01}}+2\Delta_S}\right)^2 (\Gamma_X^2-4(z-\Delta)^2) } \right)\, .\label{EffectiveGammas}
\end{equation}
\end{widetext}

Figure 6 shows the dependence of the effective relaxation times, $\tilde{\Gamma}_{I}^{-1}(z)$ (blue line) and $\tilde{\Gamma}_{D}^{-1}(z)$ (light red line), as function of applied electric field $F$. For values of field away from the central anticrossing in Fig.\ 3c, the effective relaxation time, of the eigenstates of $\tilde{H}_{\Gamma}^{(I)}(z)$, $\tilde{\Gamma}_{D}^{-1}(z)$, approaches the relaxation time of the bare effective indirect excitons, $\tilde{\Gamma}_{I}^{-1}(z)$, where the interference term (dephasing), $\tilde{\Gamma}_\Phi$, significatively diminishes. At the avoided crossing with $F_R=2.3$kV/cm, the eigenstate relaxation time reaches a minimum value of $\sim22.2$ ns. Notice, it is $ \tilde{\Gamma}_I^{-1}(z)$ which is more physically relevant, as it gives the lifetime of the qubit logical states for all values of $F$, between the initialization regime, at $F_I =  43.4$kV/cm, and the rotation regime at $F_R = 2.3$kV/cm, with corresponding values of $214$ns and $24.5$ns, in good agreement with the numerical solutions of the Lindblad equation shown on Fig.\ 4. Notice that the ratio between gate rotation and decoherence times, $U/\Gamma_I$, is monotonically dependent on the interdot distance, so that an optimal ratio is not given within the model, but is set by the QDM geometry.

\section{Concluding Remarks}
\label{conclusion}
In summary, we have shown that the exciton spectrum of a QDM can be used to define an optimally defined qubit using two spatially indirect neutral excitons. We found that the interplay of optical excitation and charge tunneling can produce optical signatures that identify the indirect exciton qubit subspace. Although the QDM is treated explicitly as an open quantum system, with exciton relaxation rates arising from spontaneous decay, the subspace of indirect qubit states has large decoherence times. This is explained by a large suppression of the interaction between the qubit subspace and the laser field. On the other hand, the qubit can be initialized with near unity fidelity via higher order couplings to the radiation field. In this manner the input state can be shelved for time intervals well beyond the direct exciton relaxation time. The use of an adiabatic bias pulse permits driving the input state into different molecular resonances, in particular, into a resonance that mixes coherently the input state with a target state of the qubit. Interestingly, a reverse bias pulse drives an adiabatic passage of the output logical state through a tunneling induced anticrossing, transferring half of its population to a spatially direct exciton. This enables the possibility of directly reading the output of the qubit rotation.
The suppression and tunability of the exciton interactions with states outside the qubit subspace contrast drastically with qubits defined via excitons with spatially direct character. This opens the possibility of using neutral exciton states as elemental blocks within a complex QDM quantum computation scheme that uses indirect excitons, such as spin qubits in molecular trions.

\acknowledgments
We thank E. Stinaff for helpful discussions, and support from NSF-DMR MWN/CIAM grant 0710581, NSF-SPIRE and the CMSS and BNNT
programs at Ohio University.


\begin{thebibliography}{99}
\bibitem{foot} Since the interdot barrier is large, we assume that the indirect excitons are not optically active, see Ref.\ [\onlinecite{Krenner}] (we set $\Omega_I =0$ for indirect exciton transitions). Inclusion of a small $\Omega_I$, or indirect exciton radiative decay, $\Gamma_{I}^{-1} \simeq 1\mu$s, does not appreciably change the results presented here.
\bibitem{numeros} Exciton transition energies and structure parameters, consistent with [\onlinecite{Bester}], and  derived from measurements [\onlinecite{tunnelings}] are as follows. QDM parameters: interdot distance $d\simeq8.4nm$, $t_{e}, t_h= 2.0, 0.1$meV, $V_F=0.08$meV, $\Omega_{T(B)}=75$meV. Exciton bare energies: $E_{_{01}^{01}X}, E_{_{01}^{10}X}, E_{_{10}^{01}X} = 1248.01, 1263.08, 1267.09$meV, $E_{_{10}^{10}X}=E_{_{01}^{01}X}+10\mu$eV. Biexciton bare energies: $E_{_{02}^{02}X}=2492.85$meV. Interband transition moments $\mu_{T(B)}=6.2e${\AA} [\onlinecite{dipoles}]. \@ Exciton-radiation couplings (controlled by the laser pulse amplitude) are in the strong pumping regime, $\Omega_{T(B)}\gg\tau_{X,(XX)}$. Exciton (biexciton) radiative recombination times $\tau_{X} (\tau_{XX})\sim 1.0\, (0.5)$ns [\onlinecite{lifetimes}].
\end{thebibliography}
\end{document}